# Critical states in superconducting plate: Structure dependence

**Shinsuke Ooi and Masaru Kato**

Department of Mathematical Sciences,
Osaka Prefecture University, 1-1,
Gakuencho, Nakamozu, Sakai, Osaka 599-8531, Japan

E-mail: sxb01036@edu.osakafu-u.ac.jp

**Abstract**. We study vortex penetration into two-layer structures of superconducting plates under a perpendicular magnetic field. We solve the heat transport equation and the Maxwell equations with the current-voltage relation for superconductor, simultaneously, and obtain magnetic flux and current densities. We show how magnetic flux structure depends on the structure, especially distance of two-layer of superconductors.

## 1. Introduction

When an external perpendicular magnetic field applied to a superconducting plate is increased, magnetic flux penetrates into the superconducting plate from its edges. Depending on a ramping rate of increased applied field, there are two kinds of vortex penetration: a critical state and a vortex avalanche [1]. If the ramping rate of the applied magnetic field is small, the critical state appears. How this critical state appears is explained as follows. When vortices penetrate into superconducting plate, the Lorentz force from the Meissner shielding current, which flow along an edge of the superconducting plate, drive vortices into the inside of the superconducting plate. Then, if there are microscopic defects in the superconductor, vortices may be pinned by these defects. This pinning force from defects acts against the driving Lorenz force. The local balance between two forces creates a metastable equilibrium state, which is the critical state. In the critical state, magnitude of the shielding current density is adjusted to the critical current density.

    On the other hand, the ramping rate is large, the vortex avalanche occurs. This phenomenon is caused by thermal fluctuation at finite temperature. The vortex, which is pinned in the pinning center, jumps to another pinning center because of the driving force and the thermal fluctuation. This movement generates heat, and temperature locally rises. Then superconductivity is weakened. Therefore, vortices further penetrate into the superconductor. Repeating this process, there appears the vortex avalanche.

    The critical states and the vortex avalanches in complicated three-dimensional structures, which are composed of several stacked superconducting strips, were investigated experimentally [2,3]. In a two-layer structure, unlike single plate case, vortices mainly penetrate from overlap regions between the two superconductors, causing some dotted or linear vortex avalanches. Like the conventional vortex avalanche, there are heat generation by vortex motion and heat conduction, and therefore shapes of avalanches depend on the distance and overlap of superconducting strips, and the size of the superconducting strip.

In this study, we investigate the critical states rather than the vortex avalanche in a two-layer structure of two superconducting plates, when the ramping rate of magnetic field is not large, and heat generation and temperature rise are small. We solve the heat transport equation and the Maxwell equation with a current-voltage relation for superconductors with three-dimensional finite element method. We obtain distribution of the magnetic flux density and the current density.

## 2. Method

We solve following three equations. The first one is the heat transport equation,

$$C \frac{\partial T}{\partial t} = \frac{\partial X}{\partial x} + \frac{\partial Y}{\partial y} + \frac{\partial Z}{\partial z} - \alpha(T - T_0) + \boldsymbol{J} \cdot \boldsymbol{E} \tag{1}$$

Here, $C$ is a specific heat, $T(x, y, z, t)$ is temperature of the superconductor, $t$ is a time, and $X(T), Y(T),$ and $Z(T)$ are the heat flow in $x, y,$ and $z$ direction, respectively. The forth term on the right hand side is the heat transfer from the superconductor to the substrate surrounding the superconductor, where $\alpha$ is a heat transfer rate, and $T_0$ is temperature of the substrate. The fifth term on the right hand side is the heat generation by the vortex creep. Here $\boldsymbol{J}$ is a current density, and $\boldsymbol{E}$ is an electric field. $X(T), Y(T),$ and $Z(T)$ are given as follows.

$$\begin{cases} X(T) = \kappa_{11} \frac{\partial T}{\partial x} + \kappa_{12} \frac{\partial T}{\partial y} + \kappa_{13} \frac{\partial T}{\partial z} \\ Y(T) = \kappa_{21} \frac{\partial T}{\partial x} + \kappa_{22} \frac{\partial T}{\partial y} + \kappa_{23} \frac{\partial T}{\partial z} \\ Z(T) = \kappa_{31} \frac{\partial T}{\partial x} + \kappa_{32} \frac{\partial T}{\partial y} + \kappa_{33} \frac{\partial T}{\partial z} \end{cases} \tag{2}$$

where $\kappa_{ij}(i, j = 1, 2, 3)$ are heat conductivity tensors.

The second equations are the Maxwell equations,

$$\begin{cases} \text{rot}(\text{rot}\boldsymbol{A} - \boldsymbol{H}_a) = \frac{4\pi}{c} \boldsymbol{J} \\ \boldsymbol{E} = -\frac{1}{c} \frac{\partial \boldsymbol{A}}{\partial t} \end{cases} \tag{3}$$

where $\boldsymbol{A}$ is a magnetic vector potential, $\boldsymbol{H}_a$ is an external magnetic field, $c$ is the speed of light. The first Maxwell equation is Ampere's law, in which we ignore displacement current, and the second Maxwell equation is Faraday's law.

The third equation is the current-voltage relation for the superconductor [4],

$$\boldsymbol{J} = \frac{\boldsymbol{E}}{\rho(\boldsymbol{J})} \tag{4}$$

Here $\rho(\boldsymbol{J})$ is a resistivity given by,

$$\rho(\boldsymbol{J}) = \begin{cases} \rho_0(\boldsymbol{J}/\boldsymbol{J}_c)^{n-1} & \boldsymbol{J} \leq \boldsymbol{J}_c, T \leq T_c \\ \rho_0 & \boldsymbol{J} > \boldsymbol{J}_c, T \leq T_c \\ \rho_n & \boldsymbol{J} > \boldsymbol{J}_c, T > T_c \end{cases} \tag{5}$$

where $n$ is a creep exponent, $\rho_0$ is a resistivity constant of superconducting state, $\rho_n$ is a resistivity of normal state, $\boldsymbol{J}_c$ is a critical current density, and $T_c$ is the critical temperature. The flux creep exponent is given by $n - 1 = n_0(1 - T/T_c)$, where $n_0$ is a constant, and it depends on the temperature. This current-voltage relation represents the fact that the resistance gradually rises due to vortices creeps as the current increases. The current and temperature dependences of the resistivity in the superconductor are shown in figure 1.

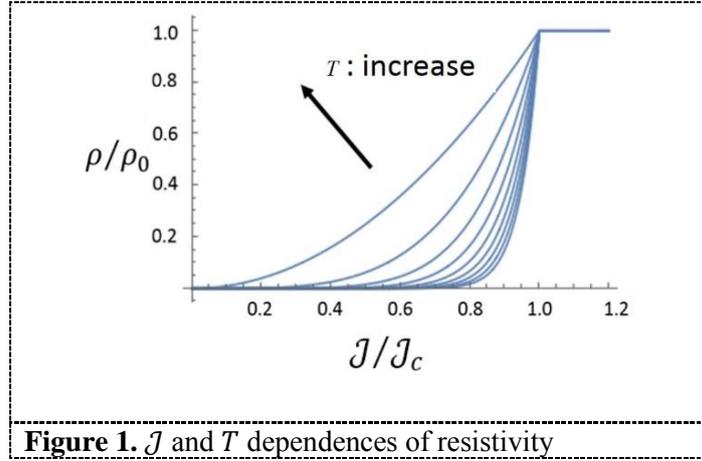

**Figure 1.** $J$ and $T$ dependences of resistivity

## 3. Results and Discussion

We consider two-layer structures of two superconducting plates, and investigate how the critical state depends on distance of two superconducting plates. The size of each superconducting plate is $30 \times 60 \times 8\ \mu m^3$. Each superconducting plate is surrounded by the substrate shown in figure 2; distance between two superconducting plates is $d_i$, and the overlap of two superconducting plates is 10 μm. The total system size is $70 \times 80 \times (64 + d_i)\ \mu m^3$. We change the distance from 0 to 56 μm.

We use material parameters corresponding to $MgB_2$, $C = 35\ kJ/K\ m^3 \times (T/T_c)^3$, $\alpha = 220\ kW/K\ m^2 \times (T/T_c)^3$, $\kappa = 0.17\ kW/K\ m \times (T/T_c)^3$, $\rho_n = \rho_0 = 7\ \mu\Omega cm$, $T_c = 39\ K$, $J_c = 3600\ kA/m^2$, $n_0 = 19$ [4]. The external applied magnetic field along $z$ axis is ramped from $H_a = 0$ kG at a constant rate, $\dot{H}_a = 3 \times 10^{-4}$ mT/s. The initial temperature of the total system is 10 K. Although the temperature of the superconductors rises, the temperature of the substrate is kept at 10 K.

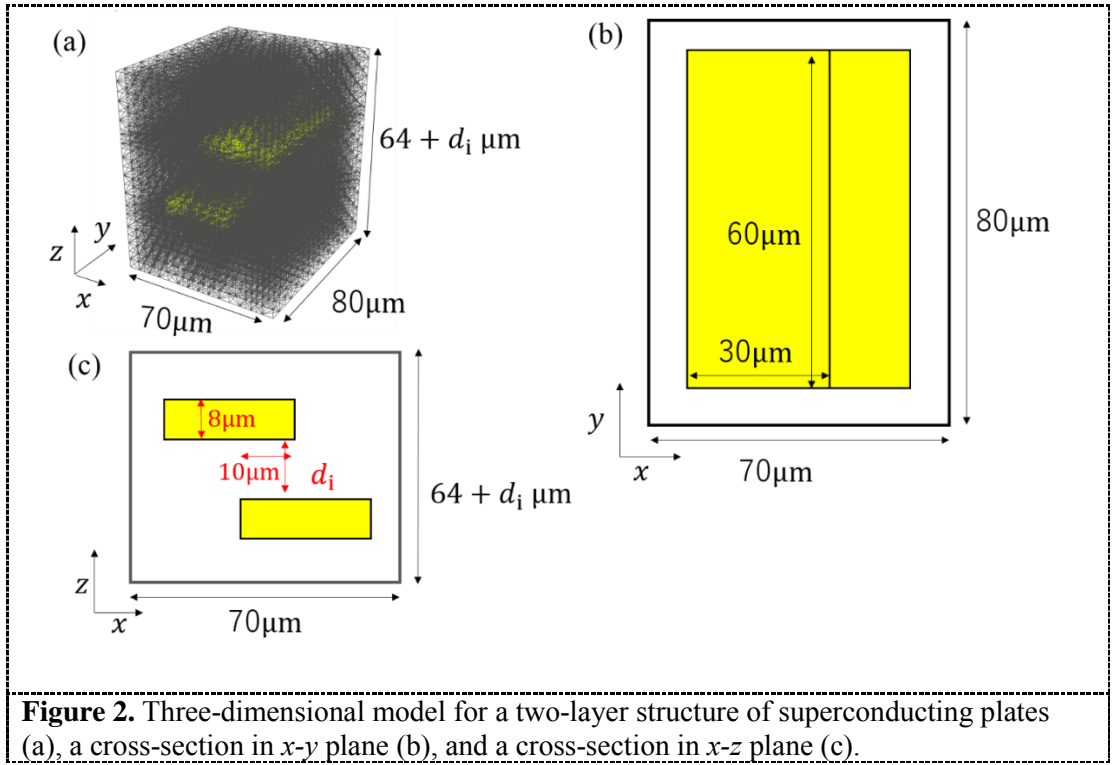

**Figure 2.** Three-dimensional model for a two-layer structure of superconducting plates (a), a cross-section in $x$-$y$ plane (b), and a cross-section in $x$-$z$ plane (c).

First, we show the critical state for $d_i = 24$ μm when the external magnetic field is $H_a = 3 \times 10^{-6}$ mT. Figures 3(a) and (b) show distribution of magnitude of magnetic flux density $B$ along $x$ direction at $y = 40$ μm, and $z = 24$ μm, 28 μm and 32 μm in the lower superconducting plate (a) and $z = 56$ μm, 60 μm and 64 μm in the upper superconducting plate (b). Figures 3(c) and (d) show distribution of the current density $\mathcal{J}_y/\mathcal{J}_c$ along $x$ direction at $y = 40$ μm, and $z = 24$ μm, 28 μm and 32 μm in the lower superconducting plate (c) and $z = 56$ μm, 60 μm and 64 μm in the upper superconducting plate (d). From figures 3(a) and (b), we can see distributions of $B$ at $z = 24$ and 64 μm have peaks at both sides of plane and distributions of $B$ at $z = 32$ and 56 μm have a peak only one side of plate and show strong asymmetry. Total distribution of $B$ is symmetric under spatial inversion about $z = 0$ plane. And current density distributions in figures 3(c) and (d) show similar properties as those of $B$. These asymmetries come from small magnetic field around the overlap regions of two superconducting plates. This is because magnetic energy becomes large if magnetic field go through the gap of two superconductors.

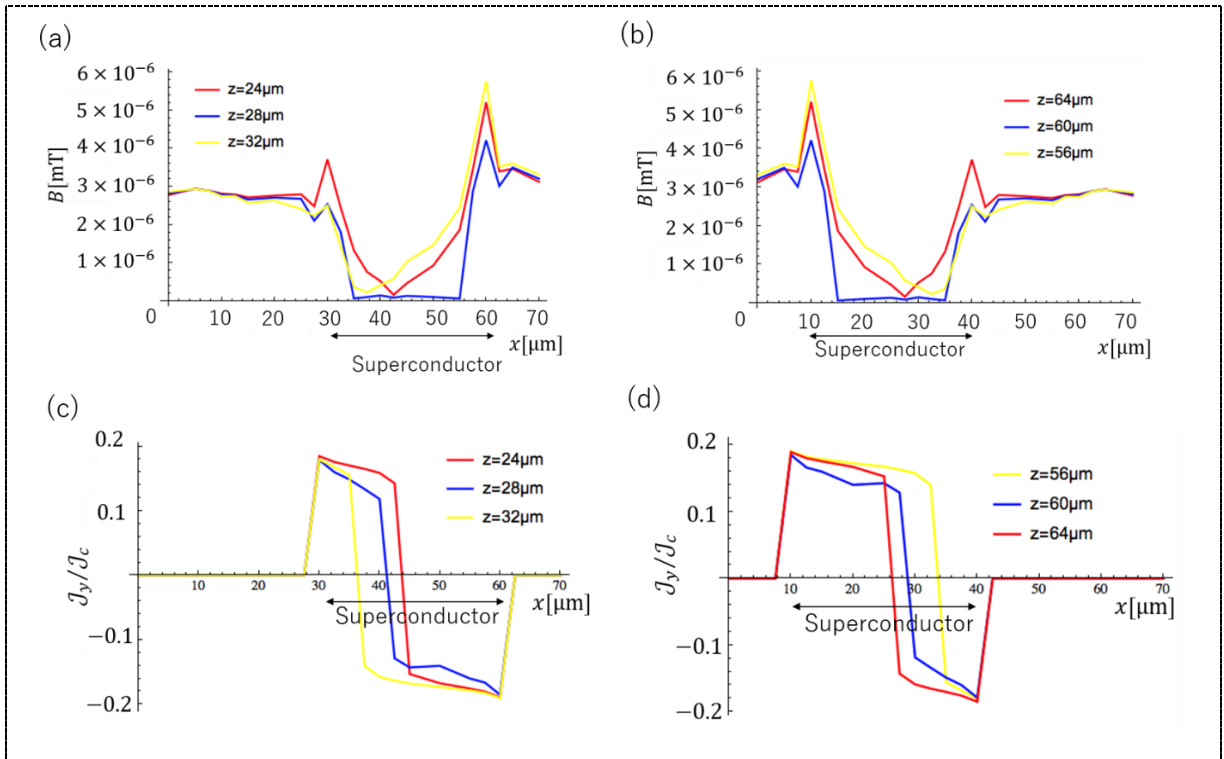

**Figure 3.** (a) and (b) are magnetic flux density and (c) and (d) are current density.

Next, we show results for changing $d_i$ from 8 to 56. Figures 4(a) and (b) show distributions of the magnetic flux density (a) and the current density (b) at $z = 32$ μm and $y = 40$ μm along $x$ direction. From figure 4(a), we can see that a peak around $x = 30$ μm appears and becomes large as increasing the distance $d_i$. Correspondingly, the minimum position of the magnetic flux density moves toward inside of the superconductor, and also distribution of the current density becomes wider as shown by an arrow in figure 4(b). In figure 5, we show the dependence of the flux density at $x = 30$ μm, $y = 40$ μm and $z = 32$ μm on the distance $d_i$. We also show max flux density $B_{max}$ at another peak at $x = 60$ μm, $y = 40$ μm and $z = 32$ μm. From this figure, we can see $B$ is saturated to $B_{max}$ when the distance is increased. Therefore, we can expect that in the limit $d_i \to \infty$, the magnetic flux density distribution becomes that of a single superconducting plate where two symmetric peaks at the edges of plate appear [5].

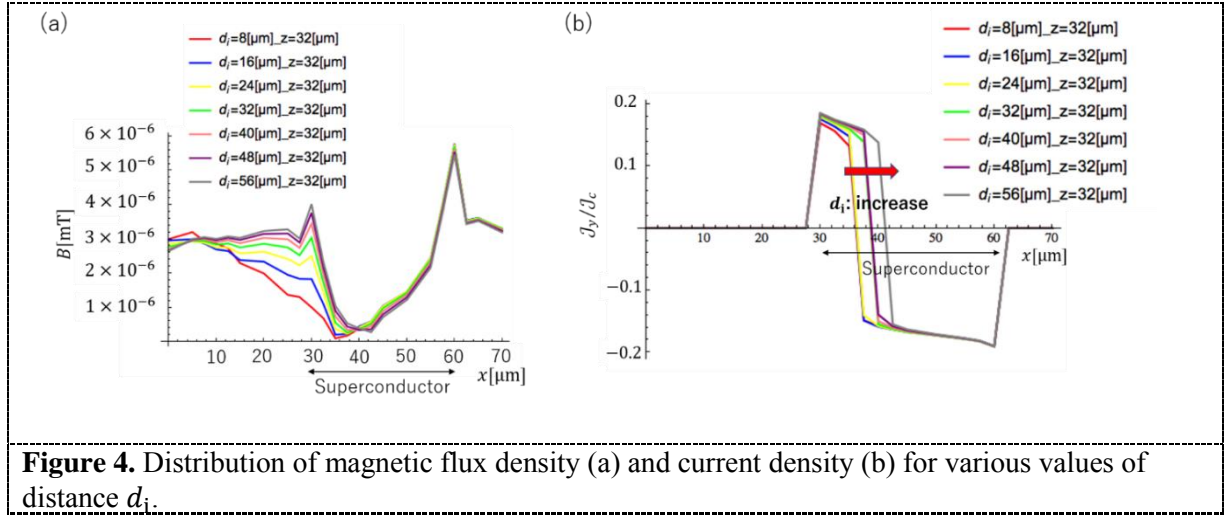

**Figure 4.** Distribution of magnetic flux density (a) and current density (b) for various values of distance $d_i$.

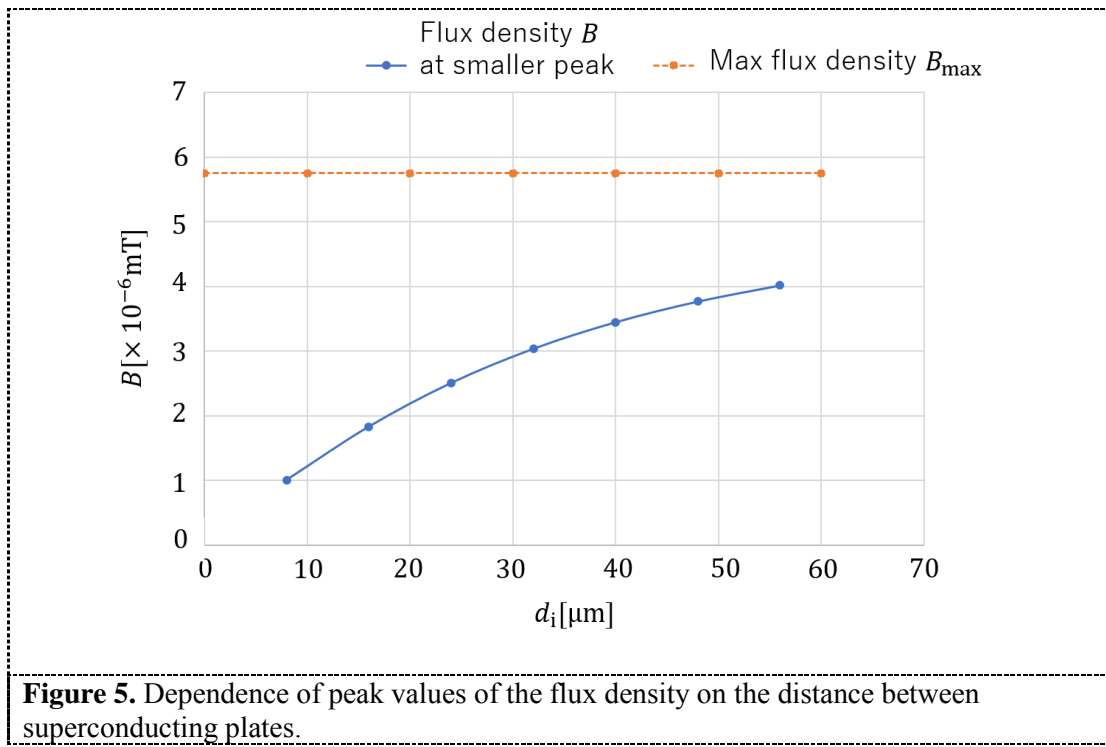

**Figure 5.** Dependence of peak values of the flux density on the distance between superconducting plates.

## 4. Summary and Future

We have studied the critical state for two-layer structure of two superconducting plates. Solving the heat transport equation and the Maxwell equations with the current-voltage relation for superconductor with three-dimensional finite element method, we have obtained distributions of the magnetic flux density and the current density. We have found asymmetric distributions of the magnetic flux density and the current density in each superconducting plate in the two-layer structure. Moreover, there is the structure dependence of the magnetic flux density. The larger the distance, the more magnetic flux penetrate from the overlapped part of two superconducting plates.

In future, we will simulate magnetic field distribution more realistically, using the boundary element method for the substrate region surrounding superconductors.

**References**
[1] Altshler E and Jhohansen T H 2004 *Rev. Mod. Phys.* **76** 471
[2] Tsuchiya Y, Mawatari Y, Ibuka J, Tada S, Pyon S, Nagasawa S, Hidaka M, Maezawa M and Tamegai T 2013 *Supercond. Sci. Technol.* **26** 095004
[3] Tamegai T, Mine A, Tsuchiya Y, Miyano S, Pyon S, Mawatari Y, Nagasawa S, and Hidaka M 2017 *Phys. C.* **533** 74-79
[4] Vestgården J I, Shantsev D V, Galperin Y M and Johansen T H 2011 *Phys. Rev. B.* **84** 054537
[5] Brandt E H 1992 *Phys. Rev. B.* **46** 8628